\documentclass[prl,aps,superscriptaddress,twocolumn,showpacs,floatfix]{revtex4-1}
\usepackage{amsfonts}
\usepackage{amssymb}
\usepackage{hyperref}
\usepackage{graphicx}
\usepackage{dcolumn}
\usepackage{bm,amsmath,verbatim}
\usepackage{mathrsfs}
\usepackage{color}

\usepackage[T1]{fontenc}
\usepackage{booktabs}

\hypersetup{colorlinks,
linkcolor=blue,          citecolor=blue,        filecolor=blue,      urlcolor=blue           }

\newcommand{\ket}[1]{|#1\rangle}
\newcommand{\bra}[1]{\langle #1 |}

\newcommand{\ii}{\text{i}}

\begin{document}

\title{Probing Complex-energy Topology via Non-Hermitian Absorption Spectroscopy in a Trapped Ion Simulator}
\author{M.-M. Cao}
\thanks{These authors contribute equally to this work}%
\affiliation{Center for Quantum Information, Institute for Interdisciplinary Information Sciences, Tsinghua University, Beijing 100084, PR China}

\author{K. Li}
\thanks{These authors contribute equally to this work}%
\affiliation{Center for Quantum Information, Institute for Interdisciplinary Information Sciences, Tsinghua University, Beijing 100084, PR China}

\author{W.-D. Zhao}
\affiliation{HYQ Co., Ltd., Beijing 100176, PR China}

\author{W.-X. Guo}
\affiliation{Center for Quantum Information, Institute for Interdisciplinary Information Sciences, Tsinghua University, Beijing 100084, PR China}

\author{B.-X. Qi}
\affiliation{Center for Quantum Information, Institute for Interdisciplinary Information Sciences, Tsinghua University, Beijing 100084, PR China}

\author{X.-Y. Chang}
\affiliation{Center for Quantum Information, Institute for Interdisciplinary Information Sciences, Tsinghua University, Beijing 100084, PR China}

\author{Z.-C. Zhou}
\affiliation{Center for Quantum Information, Institute for Interdisciplinary Information Sciences, Tsinghua University, Beijing 100084, PR China}
\affiliation{Hefei National Laboratory, Hefei 230088, PR China}

\author{Y. Xu}
\email{yongxuphy@tsinghua.edu.cn}
\affiliation{Center for Quantum Information, Institute for Interdisciplinary Information Sciences, Tsinghua University, Beijing 100084, PR China}
\affiliation{Hefei National Laboratory, Hefei 230088, PR China}

\author{L.-M. Duan}
\email{lmduan@tsinghua.edu.cn}
\affiliation{Center for Quantum Information, Institute for Interdisciplinary Information Sciences, Tsinghua University, Beijing 100084, PR China}
\affiliation{Hefei National Laboratory, Hefei 230088, PR China}

\begin{abstract}
Non-Hermitian systems generically have complex energies, which may host topological structures, such as links or knots.
While there has been great progress in experimentally engineering non-Hermitian models in quantum simulators,
it remains a significant
challenge to experimentally probe complex energies in these systems, thereby making it difficult to directly diagnose
complex-energy topology.
Here, we experimentally realize a two-band non-Hermitian model with a single trapped ion whose complex eigenenergies exhibit
the unlink, unknot or Hopf link topological structures. Based on non-Hermitian absorption spectroscopy, we couple one system level to
an auxiliary level through a laser beam and then experimentally measure the population of the ion on
the auxiliary level after a long period of time.
Complex eigenenergies are then extracted,
illustrating the unlink, unknot or Hopf link topological structure.
Our work demonstrates that complex energies can be experimentally measured in quantum simulators via
non-Hermitian absorption spectroscopy, thereby opening the door for exploring various complex-energy properties in non-Hermitian
quantum systems, such as trapped ions, cold atoms, superconducting circuits or solid-state spin systems.
\end{abstract}
\maketitle

Non-Hermitian physics has witnessed a rapid development
in recent years due to the finding of
peculiar topological properties without Hermitian counterparts~\cite{ChristodoulidesNPReview,XuReview,ZhuReview,UedaReview,BergholtzReview}. For instance, non-Hermitian systems can have complex
energy spectra with exceptional points or rings~\cite{Zhen2015nat,Xu2017PRL,Nori2017PRL,Kozii2017,Zyuzin2018PRB,Zhou2018,Cerjan2018PRB,Yoshida2018PRB,Zhao2018PRB,Carlstrom2018PRA,
HuPRB2019,Wang2019PRB,Yoshida2019PRB,Ozdemir2019,Cerjan2019nat,Kawabata2019PRL,Zhang2019PRL,Zhang2020PRL,
Chuanwei2020PRL,Yang2020PRL,Wang2021PRL,Nagai2020PRL,Nori2021PRL,Yuliang2021,QiuCW2022PNAS,Cheng2022PRL}. This is very different from Hermitian
systems whose energies are always real. Such complex energies may also exhibit loop structures in the complex-energy plane
with a nonzero winding number, which is closely related to non-Hermitian skin effects~\cite{Yao2018PRL1,TonyLee,Xiong2018JPC,Torres2018PRB,Kunst2018PRL,Okuma2020PRL,Slager2020PRL,ChenFang2020PRL}. In addition, more complex
knotted topological structures can arise in complex energies, leading to topological complex-energy braiding~\cite{Fan2020PRB, Zhao2021PRL, Mong2021PRB,Wojcik2021}.
Remarkably, such complex-energy topology
has recently been experimentally observed in two coupled ring resonators~\cite{Fan2021Nature} and {acoustic metamaterials~\cite{Qiu2023PRL}}.

\begin{figure}[t]
	\includegraphics[width=3.4in]{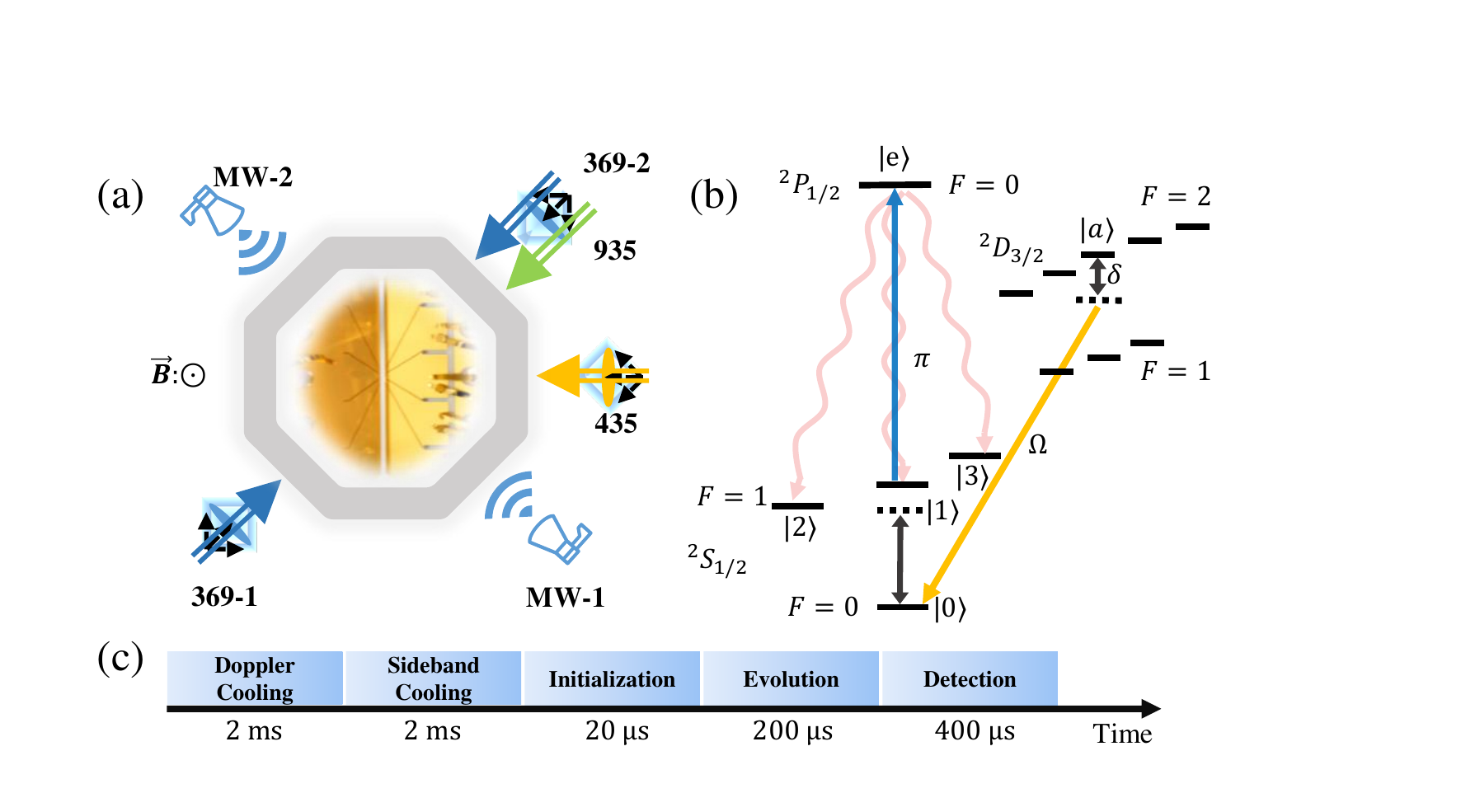}
	\caption{(a) Schematic of our experimental setup where
		a $^{171}$Yb$^{+}$ ion is confined in a linear Paul trap.
		We use two 369 nm laser beams to perform cooling, pumping and detection of an ion and to
		produce a population loss, respectively.
		A 935 nm laser beam is applied to repump the leakage back to the Doppler cooling cycle.
		We also shine a focused 435 nm narrow-band laser beam to drive the transition between the $\ket{0}$ level and the auxiliary level $\ket{a}$.
		Microwaves are used to implement the Hermitian part of the models in Eq.~(\ref{NHRMsysHam}) and Eq.~(\ref{LinkHam}).
		(b) Schematic of laser and microwave configurations.	
		Colored and black arrows specify the transitions induced by laser beams shown in (a) and microwave fields, respectively.
		Note that energy levels are plotted for visual clarity rather than based on a real energy scale.
		(c) Experimental sequences consisting of Doppler cooling, sideband cooling, initialization, time evolution and detection.
	}
	\label{fig1}
\end{figure}

Quantum simulators, such as trapped ions, cold atoms, superconducting circuits or solid-state spin systems,
provide flexible and powerful platforms to perform quantum simulations~\cite{ColdAtomReview,SolidStateSpinReview,SuperconductingCircuitsReview,QuantumSimulationReview,TrappedIonReview}.
Recently, creating non-Hermitian Hamiltonians in these simulators has become an important research goal due to
potential quantum information processing applications~\cite{Murch2019NP}.
With experimental realizations of non-Hermiticity in these quantum systems,
the parity-time ($\mathcal{PT}$) symmetry breaking has been observed~\cite{Du2019Science,LuoNC2019,Murch2019NP,Mottonen2019PRB,WeiZhang2021PRL,ChenCW2021PRA,Du2021PRL,Deng2021PRL,Jo2022NP}.
More recently, the diffusion map method is employed to learn distinct knotted topological phases based on experimental raw data
in a solid-state spin non-Hermitian system~\cite{Duan2021arxiv}.
However, the deciding feature of knotted structures in
complex energies has not been probed. In fact, it remains a difficult task in these quantum systems to experimentally measure
complex eigenenergies of a non-Hermitian system. Very recently, a protocol (dubbed non-Hermitian absorption spectroscopy)
has been proposed
to measure complex energy spectra in these quantum simulators~\cite{Xu2022PRL}, providing an opportunity to probe complex-energy
topology in a trapped ion simulator.

\begin{figure}[t]
\includegraphics[width=3.4in]{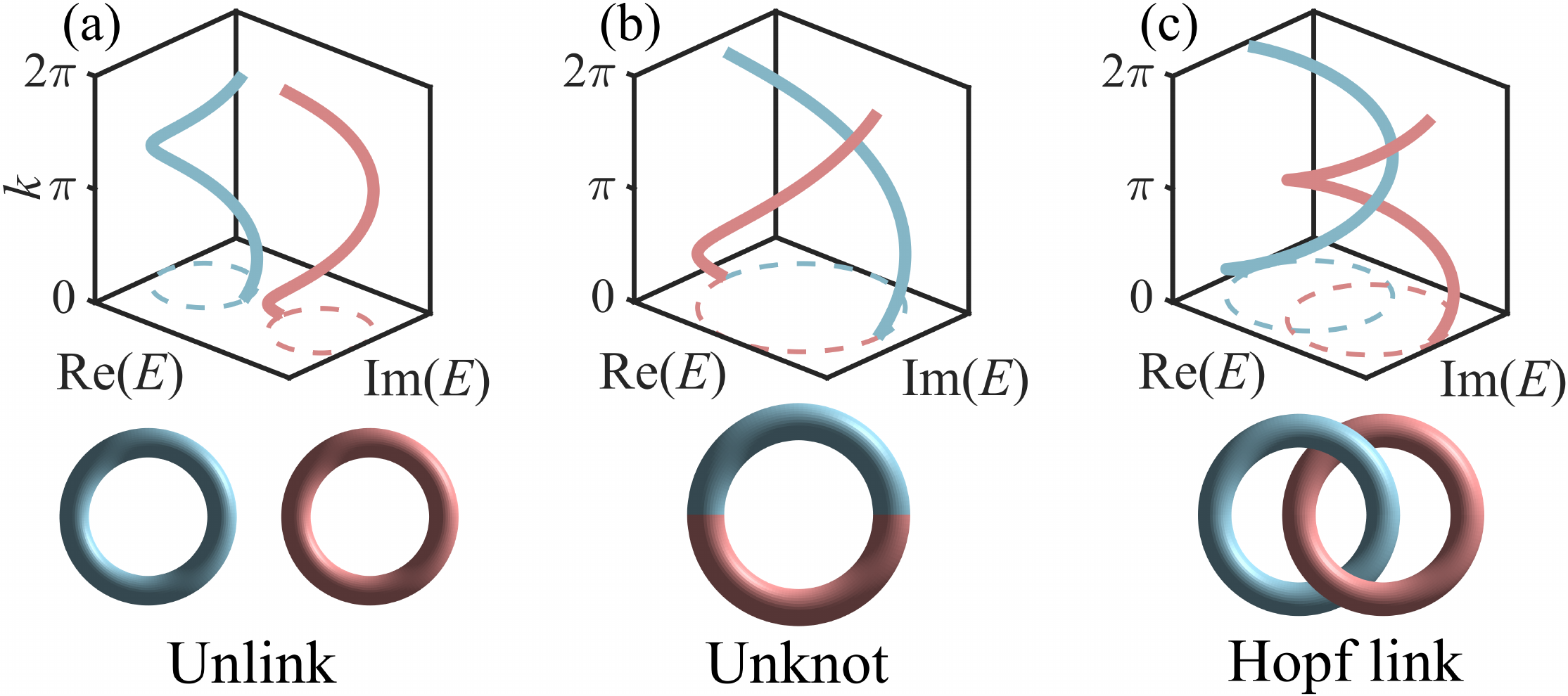}
\caption{Schematics of the (a) unlink, (b) unknot and (c) Hopf link topological structures
formed by complex energies in the $[\text{Re(E)}, \text{Im(E)}, k]$ space by connecting the energies at $k=0$ and $k=2\pi$.
}
\label{fig1.5}
\end{figure}

Here, we experimentally realize a non-Hermitian model, which hosts knotted complex energies,
using a single ${}^{171}$Yb$^{+}$ ion in a Paul trap.
The Hermitian part of the model is realized through microwave pulses driving the transition between two
system levels, and
the non-Hermitian part is implemented through a resonant laser
beam driving an ion from
one system level $|1\rangle$ to a $P_{1/2}$ level, leading to a population loss of an ion via the spontaneous emission (as sketched in Fig.~\ref{fig1}).
To experimentally measure their complex energies based on non-Hermitian absorption spectroscopy,
we prepare an ion on a $^{2}D_{3/2}$ level (serving as an auxiliary level) and then
shine a weak laser beam to couple it to the system level $|0\rangle$.
After a long period of time, we measure the probability of the ion on the auxiliary level.
Complex energies are finally extracted by fitting the measured probability with respect to a detuning.
The measured complex energies exhibit the unlink, unknot or Hopf link topological structures {(schematics shown in Fig.~\ref{fig1.5})}, which agree
well with theoretical results.

We start by considering a {modified} non-Hermitian Rice-Mele model described by
\begin{equation}
	\label{NHRMsysHam}
	H_{\text{MRM}}(k) =(J_0\ket{0}\bra{1}+\text{H.c.}) - 2(J_z + \ii \gamma) \ket{1} \bra{1},
\end{equation}
where $J_0=\sqrt{J_x^2+J_y^2}$ with $J_x=J_1 + J_2 \cos k$, $J_y = J_2 \sin k$, and $J_z = J_3 \sin k + m_z$. Here,
$J_1$, $J_2$, $J_3$ and $m_z$ are real system parameters, and $k$ is the momentum taken in the interval $[0,2\pi]$.
$|0\rangle$ and $|1\rangle$ denote two system levels encoded in two hyperfine states of an ion,
and $\gamma$ represents the decay strength of an ion on the $|1\rangle$ level.
{Note that this model has the same eigenenergy as the non-Hermitian Rice-Mele model $H_{\text{RM}}(k) =(J\ket{0}\bra{1}+\text{H.c.}) - 2(J_z + \ii \gamma) \ket{1} \bra{1}$  where $J=J_x-\ii J_y$~\cite{Yi2020PRL}.}
Due to the existence of the loss term, the system's eigenenergies 
$
E_{\pm}(k)=\pm \sqrt{J_0^2+(J_z+\ii \gamma)^2}-(J_z+\ii \gamma)
$
  are complex,
giving rise to complex-energy topology.
For instance, when $J_3 \neq 0$, the complex eigenenergies may exhibit the unlink topological structure consisting of two loops in the complex-energy plane with
the winding number of $1$ [see Fig.~\ref{fig2}(b1)].
For other parameter values, the model can host the unknot topological structure with one loop formed by two connected bands characterized
by a half winding number [see Fig.~\ref{fig3}(a) and (b)].

\begin{figure*}[t]
	\centering
	\includegraphics[width=\textwidth]{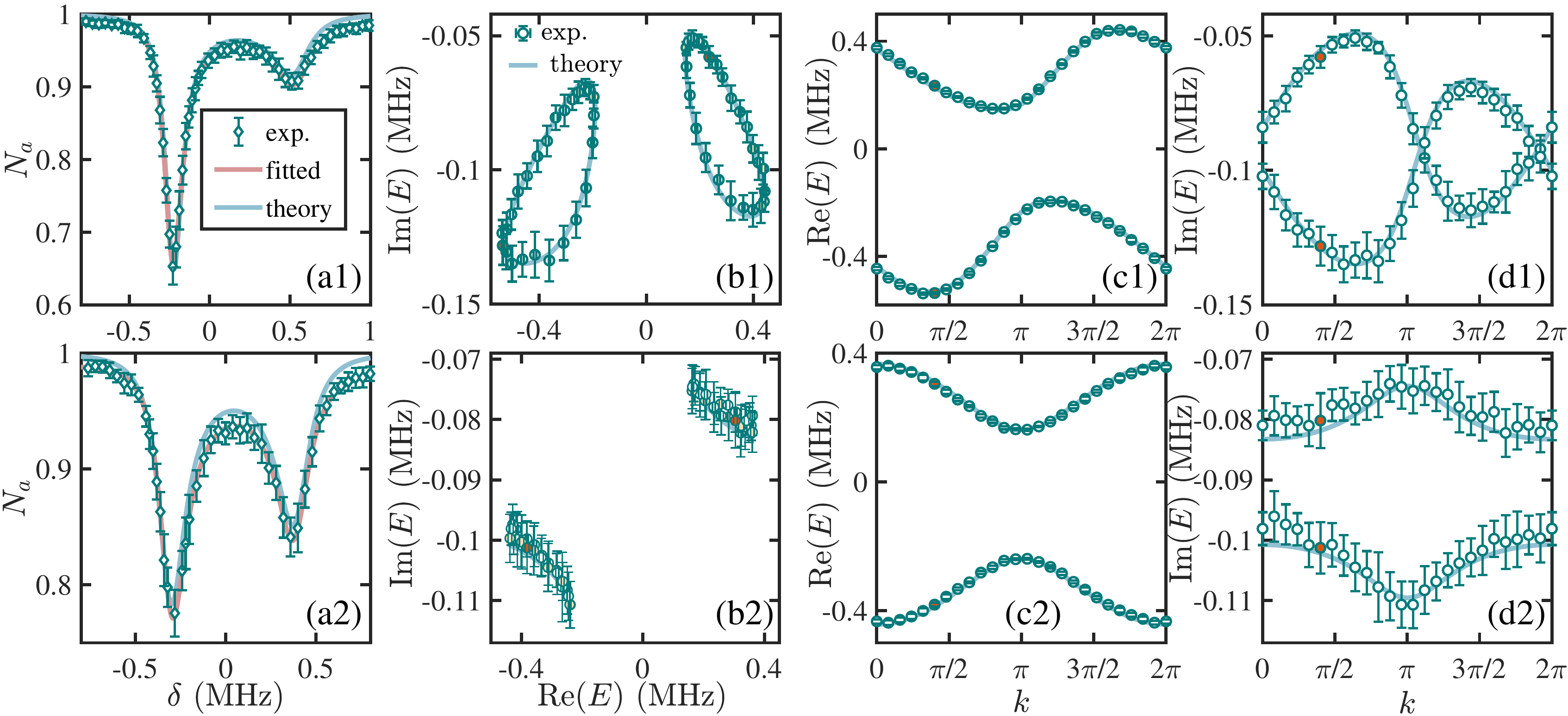}
	\caption{Experimentally measured complex eigenenergies of the {modified} non-Hermitian Rice-Mele model in Eq.~(\ref{NHRMsysHam}).
		(a1),(a2) Spectral lines of $N_a$ with respect to the detuning $\delta$ obtained by experimental measurements (diamonds with error bars), fitting the experimental data (red lines) and numerical simulations (green lines) at $k=2\pi/5$ [the corresponding extracted energies are highlighted by red circles in (b)--(d)].
		In the numerical simulation, we set $N_0 = 1$.
		 (b1),(b2) Complex energies in the complex-energy plane.
		Real (c1),(c2) and imaginary parts (d1),(d2) of complex energies as a function of the momentum parameter $k$.
		The energies are extracted by fitting spectral lines (circles with error bars) or
		 obtained by diagonalizing the system Hamiltonian (solid lines) based on the realized system parameters.
	    Here, we realize a topologically nontrivial Hamiltonian with $J_1=0.315$ MHz, $J_2=0.098$ MHz, $J_3=0.122$ MHz, $m_z=0.035$ MHz, $\gamma=0.092$ MHz
	    and $\Omega=0.019$ MHz in (a1)--(d1) and a trivial Hamiltonian with $J_3=0$ and $m_z=0.038$ MHz in (a2)--(d2) [the other parameters are the same as those in (a1)--(d1)].
	    In both Fig.~\ref{fig2} and Fig.~\ref{fig3},
	    experimental data are averaged over $20$ experimental repetitions (each contains $1000$ shots) with error bars estimated by
	    the standard deviation of the $20$ rounds of experiments.
	}
	\label{fig2}
\end{figure*}

We experimentally implement the Hamiltonian in Eq.~(\ref{NHRMsysHam}) with a single ${}^{171}$Yb$^{+}$ ion in a Paul trap as sketched in Fig.~\ref{fig1}.
The two states $\ket{0}$ and $\ket{1}$ are encoded in two hyperfine states $|F=0, m_F=0 \rangle$ and $|F=1, m_F=0 \rangle$ in the $\text{S}_{1/2}$ ground-state manifold, respectively. The hyperfine energy splitting between the two states is $\omega_\text{HF}\approx 2\pi \times 12.6$ GHz (here and henceforth we set $\hbar = 1$).
Two microwaves are applied to generate the coupling between $\ket{0}$ and $\ket{1}$ {so that $J_0$ and $J_z$ can be tuned by controlling the intensity and detuning of microwaves, respectively}. 
To create a population loss
of an ion on the $\ket{1}$ level, we shine a resonant 369 nm laser beam on the ion
to excite it from the $|1\rangle$ level to the upper $^2 \text{P}_{1/2}$ state [see the blue arrow in Fig.~\ref{fig1}(b)].
The laser beam is $\pi$-polarized so that excitations from other Zeeman levels ($\ket{^2\text{S}_{1/2}, F=1, m_F=\pm1}$) are forbidden by selection rules.
Since the $^2 \text{P}_{1/2}$ state has a short lifetime ($\tau \approx 8.12$ ns)~\cite{Monroe2009PRA}, it will spontaneously decay to all the three Zeeman levels [denoted by wave arrows in Fig.~\ref{fig1}(b)], rendering a large decay rate on the $^2 \text{P}_{1/2}$ state. 

At this stage, the dynamics of a state is described by the following master equation
\begin{equation}
	d \rho / dt = -i( H_\text{eff} \rho - \rho H_\text{eff}^\dagger) + \sum\nolimits_{\mu=1}^{3} 2 L_\mu \rho L_\mu^\dagger,
\end{equation}
where $\rho$ is the density matrix, $L_\mu = \sqrt{\Gamma_\mu} | \mu \rangle   \langle e |$ ($\mu = 1, 2, 3$) are Lindblad operators, $H_\text{eff} = H_h - i \sum_{\mu=1}^{3} L_\mu^\dagger L_\mu = H_h - i \Gamma_e | e \rangle \langle e |$ with $\Gamma_e = \sum_{\mu=1}^{3} \Gamma_\mu$ and $H_h$ being the
Hermitian Hamiltonian realized in our system [see Fig.~\ref{fig1}(b) for state information].
For the master equation, one can adiabatically eliminate the $^2 \text{P}_{1/2}$ state to obtain an effective Hamiltonian described by
Eq.~(\ref{NHRMsysHam}) where the $\ket{1}$ level experiences a decay; the decay rate $\gamma$ can be varied by tuning the laser power~\cite{SM}.

In experiments, we initially prepare an ion on the $\ket{0}$ level via Doppler cooling, sideband cooling and optical pumping.
A resonant 435 nm laser beam is then
applied to excite the ion to the $|a\rangle=|^2 \text{D}_{3/2}, F=2, m_F=0\rangle$ level
through quadrupole transitions so as to prepare the ion on the auxiliary level.
Note that the auxiliary level $\ket{a}$ has a long lifetime (up to $52.7$ ms), allowing for long time evolution without decay~\cite{osti_1595883}.
After that, we immediately turn on microwave pulses and the 369 nm laser beam which
implement the Hermitian and non-Hermitian part of the Hamiltonian, respectively.
At the same time, the power of the 435 nm laser beam is turned down to ensure that the auxiliary level is weakly coupled to
the system level $\ket{0}$ (the Rabi frequency is $\Omega$), and the detuning $\delta$ is adjusted to a fixed value.
The system then evolves for $t=200\,\mu\text{s}$ under the full Hamiltonian
$
	H_f(k) = H_{\text{MRM}}(k) + {\Omega}/{2} (\ket{0}\bra{a} + \ket{a}\bra{0}) - \delta \ket{a} \bra{a}
$. 
At the end of the evolution, the population on the auxiliary level is given by
\begin{equation}
	\label{NaTheory}
	N_a(t) = N_0 \langle a|e^{-i H_f t}|a\rangle,
\end{equation}
where $N_0$ is the initial population on the auxiliary level.
Experimentally, we perform the detection of the population $N_s$ of an ion in the $S_{1/2}$ manifold
to determine the population on the auxiliary level through $N_a=1-N_s$
[see Fig.~\ref{fig1}(c) for experimental sequence and Supplemental Material~\cite{SM} for more experimental details].

The experimentally measured spectral lines ($N_a$ versus $\delta$) enables us to
extract both real and imaginary parts of eigenenergies of our realized non-Hermitian model.
Specifically,
since the Hamiltonian in Eq.~(\ref{NHRMsysHam}) is in a generic form involving all possible terms in our experimental setup,
{its system parameters
including $J_0$, $J_z$ and $\gamma$ can be extracted by fitting the measured spectral lines based on Eq.~(\ref{NaTheory}). 
This is done by finding a set of parameters under which the results from Eq.~(\ref{NaTheory}) fit the experimental data best.}
Once we obtain them, the complex eigenenergies are immediately {calculated}.
In a real experiment, we also need to consider the initial population $N_0$ on the auxiliary level as a fitting parameter because
it is usually smaller than one due to the initial preparation error
caused by residual phonon and laser dephasing.

Figure~\ref{fig2}(a1) and (a2) display two typical experimentally measured spectral lines.
We find that the experimental results are very well characterized by a fitted line, giving rise to complex energies,
which are in good agreement with the theoretical results [as shown by red circles in Fig.~\ref{fig2}(b)--(d)].
In Fig.~\ref{fig2}(a1) and (a2), one may also notice some slight discrepancies of the fitted line from the theoretical results.
This is attributed to the fact that $N_0$ is slightly smaller than one in a real experiment (as reflected by the experimental results at a large detuning),
while it is set to one (an ideal value) in the theoretical simulation.

{By tuning the system parameters $J_0$ and $J_z$ which depend on $k$}, we experimentally measured all complex energies for all momentum parameter $k$ and
plotted them in Fig.~\ref{fig2}(b)--(d).
We see that the measured energy spectra agree well with the theoretical results, demonstrating the feasibility of our method in detecting the
complex energies of non-Hermitian quantum systems.
More interestingly, the measured complex energies clearly capture the unlink and trivial structures for the topologically nontrivial
and trivial phases, respectively.
For the nontrivial phase, the loop structure in the complex-energy plane can be characterized by the winding number~\cite{Ueda2018PRX}
\begin{equation}
	\label{windingNumber}
	w_n = \int_{0}^{2\pi} \frac{d k}{2 \pi} \partial_k \arg [E_n(k)-E_{B}],
\end{equation}
which counts the number of times that the $n$th energy band winds around a base energy $E_{B}$ in the complex-energy plane.
For a topologically nontrivial system [Fig.~\ref{fig2}(b1)--(d1)], both energy bands form closed loops, and the winding number is given by $w_n=1$ for $n=1$ and $2$ for properly chosen base energies $E_{B}$. However, for a trivial system [Fig.~\ref{fig2}(b2)--(d2)], the energy bands do not exhibit topological properties, and the winding number is zero for any $E_{B}$.

Figure~\ref{fig2}(c1) and (c2) also illustrate that one can accurately probe the real parts of complex energies with very small errors.
However, for the imaginary parts displayed in Fig.~\ref{fig2}(d1) and (d2), we observe much larger errors; these errors are also related to the magnitude of imaginary eigenenergies, that is, the errors become larger as $|\text{Im}(E)|$ increases.
Such phenomena about errors might be attributed to the structure of spectral lines.
As proved in~\cite{Xu2022PRL}, the spectral line is a combination of multiple absorption dips, with the center position and half width of each dip
closely related to the real and imaginary part of an eigenenergy, respectively.
Due to experimental imperfections, the probed spectral lines have noises as shown in Fig.~\ref{fig2}(a1) and (a2), which leads to finite errors in the detection of complex energies.
A detailed analysis about errors~\cite{SM} shows that 
the center positions of absorption dips are robust against experimental noises, while the half widths are not, so that the errors in the real parts of eigenenergies are smaller,
in contrast to those in the imaginary parts.
Furthermore, as $|\text{Im}(E)|$ increases, the absorption dip becomes wider and shallower making its half width more sensitive to experimental noises, 
which may explain the observed phenomena about errors.
One can also find the results for a shorter evolution time $t=80$ $\mu$s in Supplemental Material~\cite{SM}.

\begin{figure}[t]
	\centering
	\includegraphics[width=3.4in]{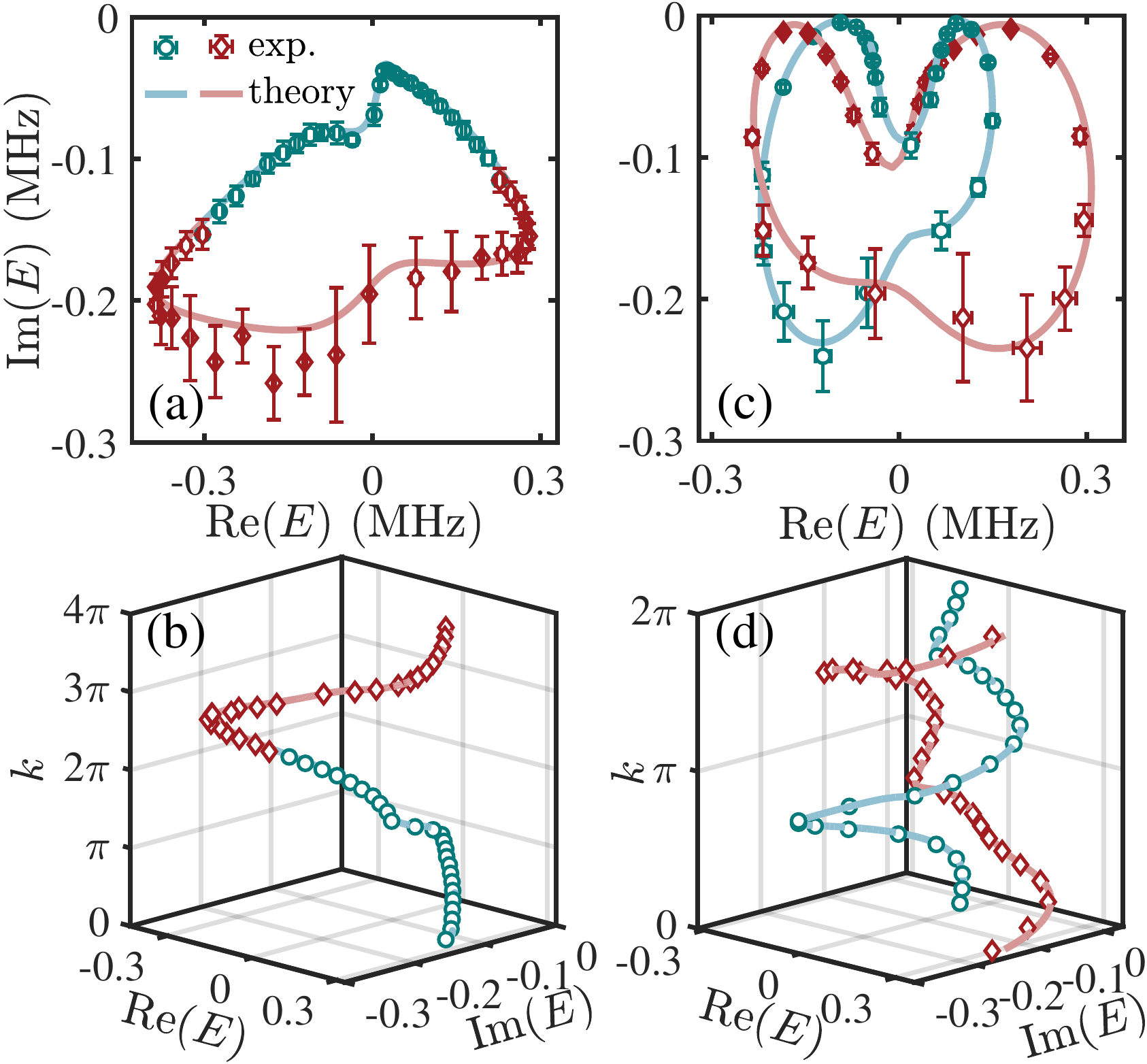}
	\caption{
		Experimentally measured complex energies with the unknot (a),(b) and Hopf link (c),(d) topological structures.
		Complex energies are plotted in the complex-energy plane in (a) and (c), and in the $[\text{Re}(E), \text{Im}(E), k]$-space
		in (b) and (d).
		We realize the model in Eq.~(\ref{NHRMsysHam}) with $J_1=0.195$ MHz, $J_2=0.098$ MHz, $J_3=0.100$ MHz, $m_z=0.038$ MHz and $\gamma=0.127$ MHz,
		in (a) and (b) and the model in Eq.~(\ref{LinkHam}) with $m_x=0.13$ MHz, $g_1=0.05$ MHz, $g_2=0.08$ MHz, $\gamma_0=0.15$ MHz and $g_3=0.07$ MHz
		in (c) and (d).
		In (b), we shift the second band (in red) along the $z$-axis by $2\pi$ in order to display the full periodicity of the energy band.
	}
	\label{fig3}
\end{figure}

For the modified non-Hermitian Rice-Mele model,
the two energy bands become closer and finally merge with each other
as we vary system parameters, e.g., decrease $J_1$, leading to one connected loop in the complex-energy plane as shown in Fig.~\ref{fig3}(a).
Such a structure is also referred to as the unknot. Our experimentally measured complex energies, which are in good agreement with the theoretical results,
also illustrate the existence of the unknot topological structure [see Fig.~\ref{fig3}(a)].
The structure of the two bands can be further demonstrated by plotting the energy bands in the $[\text{Re}(E), \text{Im}(E), k]$-space, where the first band (in green) and the shifted second band (in red) form a smooth curve from $k=0$ to $k=4\pi$.
Such a feature can be described by a modified winding number defined as~\cite{Qibo2020}
\begin{equation}
	\label{windingNumber2}
	W_n = \int_{0}^{2 m \pi} \frac{d k}{2 m \pi} \partial_k \arg [E_n(k)-E_{B}],
\end{equation}
where $m$ is the smallest positive integer satisfying $E_n(k+2m\pi)=E_n(k)$.
For the complex energies in Fig.~\ref{fig3}(a) and (b), the winding number measured relative to a base energy $E_B$ inside the loop is given by $W_n = 1/2$,
characterizing the $4\pi$ periodicity of each energy band. The experimental results in Fig.~\ref{fig3}(b) well characterize this feature.

{We now extend our scheme to detect more complex topological structures in complex energies by
considering another two-band Hamiltonian~\cite{Fan2021Nature}
\begin{equation}
\label{LinkHam}
H_{\text{LK}}(k) = m_x \sigma_x + g \ket{1} \bra{1},
\end{equation}
where $g=2[ g_1 \cos k  +  g_2 \cos 2k  +  \ii (g_3 \sin 2k -  {\gamma_0}/{2}) ]$ with $m_x$, $g_1$, $g_2$, $g_3$ and $\gamma_0$ being real parameters.
The eigenenergies of this Hamiltonian can exhibit the Hopf link structure for some parameters values [see Fig.~\ref{fig3}(d)]. 
Similarly, we experimentally implement this Hamiltonian in our trapped ion platform and probe its complex energies based on non-Hermitian absorption spectroscopy.}

In Fig.~\ref{fig3}(c) and (d), we show our experimentally measured complex energies as well as the theoretical results,
illustrating that each energy band forms a closed loop in the complex-energy plane.
In contrast to the unknot structure shown in Fig.~\ref{fig3}(a), the two bands do not merge even if they are close to each other.
In fact, except circling around a base energy, the energy bands also wind around each other from $k=0$ to $k=2\pi$, forming a Hopf link structure
characterized by the winding number~\cite{Fan2021Nature,Fu2018PRL}
\begin{equation}
	\nu_{mn} = \int_{0}^{2\pi} \frac{dk}{\pi} \partial_k  \arg  [E_m (k) - E_n (k)],
\end{equation}
with $\nu = \nu_{12} \in \mathbb{Z}$ being twice the number of times that the two bands wind around each other.
The calculated winding number based on experimental data is given by $\nu = 2$, which agrees with the fact that each band winds
around the other band once as shown in Fig.~\ref{fig3}(d).
In addition, we compute $\nu$ using the experimental data in Fig.~\ref{fig3}(a), giving $\nu = 1$ in agreement with the half winding of each energy band with respect to the other one.
The topology of separable bands can also be described by conjugacy classes of braid groups which have a one-to-one correspondence with knots or links~\cite{Zhao2021PRL}.
Thus, the topological invariant $\nu$ can be interpreted as the degree of the braid, i.e., the number of times that the two bands braid with each other when projected onto the $[\text{Re}(E), k]$-plane.

In summary, we have experimentally diagnosed the knotted topology in complex energies by probing 
both real and imaginary parts of complex energies of a non-Hermitian model with a single trapped ion. 
The experimental results illustrate the unlink, unknot or
Hopf link topological structures in complex energies, which agree well with theoretical results. 
{Such a method can be straightforwardly extended to other quantum simulators, such as cold atoms, superconducting circuits or
solid-state spin systems.}
Our experiment thus provides a basis and opens an avenue for exploring various 
complex-energy properties in non-Hermitian quantum systems.


\begin{acknowledgments}
This work was supported by the Innovation Program for Quantum Science and Technology (Grant No. 2021ZD0301601), the Tsinghua University Initiative Scientific Research Program, and the Ministry of Education of China. Y.X. acknowledges in addition support from the National Natural Science Foundation of China (Grant No. 11974201) and Tsinghua University Dushi Program.
\end{acknowledgments}

\newpage

\appendix

\begin{widetext}

In the Supplemental Material, we will
derive the effective non-Hermitian Hamiltonian based on the master equation in Section S-1,
show the experimental details about our system in Section S-2,
provide detailed analysis about errors in our experiments in Section S-3,
and finally demonstrate the feasibility of detecting the complex energy spectra using a shorter evolution time in Section S-4.

\section{S-1. Derivation of the Effective Hamiltonian}
In this section, we will derive the effective Hamiltonian $H_f$ in the main text based on the master equation.
We start with the entire system including the system levels $\ket{0}$ and $\ket{1}$, the other two Zeeman states $\ket{2}$ and $\ket{3}$, the excited state $\ket{e}$ and the auxiliary level $\ket{a}$. 
The dynamics of the entire system is governed by the master equation ($\hbar = 1$)
\begin{equation}
	\label{fullME}
	\begin{aligned}
		\frac{d \rho}{dt} &= -i [H,\rho] + \sum_{\mu=1}^{3} (2 L_\mu \rho L_\mu^\dagger - \{ L_\mu^\dagger L_\mu, \rho \} )
		\\& = -i H_\text{eff} \rho + i \rho H_\text{eff}^\dagger + \sum_{\mu=1}^{3} 2 L_\mu \rho L_\mu^\dagger.
	\end{aligned}
\end{equation}
Here $H = H_{h} + H_{c} + J_L (\ket{1} \bra{e} + \ket{e} \bra{1})$ with $H_{h} = [(J_x - i J_y) \ket{0}\bra{1} + \text{H.c.}] - 2 J_z \ket{1} \bra{1}$ being the Hermitian part of the system Hamiltonian (for generality, we here also consider the $\sigma_y$ term), $H_c = \frac{\Omega}{2} (\ket{0} \bra{a} + \ket{a} \bra{0}) - \delta \ket{a} \bra{a}$ denoting the coupling between system and auxiliary levels, and $J_L$ denoting the coupling strength between the $\ket{1}$ level and an excited $P_{1/2}$ state $\ket{e}$, 
the Lindblad operators are defined as $L_\mu = \sqrt{\Gamma_\mu} \ket{\mu} \bra{e}$ for $\mu = 1,2,3$,
and the effective Hamiltonian is given by $H_\text{eff} = H - i \sum_{\mu=1}^{3} L_\mu^\dagger L_\mu = H - i \Gamma_e \ket{e} \bra{e}$ with $\Gamma_e = \sum_{\mu=1}^{3} \Gamma_\mu = 1/\tau_e$, $\tau_e$ being the lifetime of the $\ket{e}$ level.
For the $^{171} \text{Yb}^{+}$ ion, we have $\tau_e \approx 8.12$ ns and $\Gamma_e \approx 123$ MHz~\cite{yb171plusData}.
We assume that $\Gamma_e$ is much larger than $J_L$ so that $\rho_{ei} = \bra{e} \rho \ket{i} \sim 0$ and $\partial_t \rho_{ei} \sim 0$ for any $\ket{i}$, thus we may adiabatically eliminate the $\ket{e}$ level to obtain the dynamics for the rest of the system.

Specifically, we first write down all the entries of $\rho$ that involve the $\ket{e}$ level, given by
\begin{align}
	\partial_t \rho_{e0} &= -i J_L \rho_{10} - \Gamma_e \rho_{e0} + i (J_x + i J_y) \rho_{e1} + \frac{i\Omega}{2} \rho_{ea},
	\\ \partial_t \rho_{e1} &= -i J_L \rho_{11} + i(J_x - i J_y) \rho_{e0} - (\Gamma_e + 2 i J_z) \rho_{e1} + i J_z \rho_{ee},
	\\ \partial_t \rho_{e2} &= 0, \label{rhoe2}
	\\ \partial_t \rho_{e3} &= 0, \label{rhoe3}
	\\ \partial_t \rho_{ea} &= -i J_L \rho_{1a} + \frac{i\Omega}{2} \rho_{e0} - (\Gamma_e + i \delta) \rho_{ea},
	\\ \partial_t \rho_{ee} &= -2 \Gamma_e \rho_{ee} - i J_L \rho_{1e} + i J_L \rho_{e1}.
\end{align}
Since $\rho(t=0) = \ket{a} \bra{a}$, by Eqs.~(\ref{rhoe2}) and (\ref{rhoe3}) we have $\rho_{e2}(t) = 0$ and $\rho_{e3}(t) = 0$.
Then we apply the adiabatic elimination by assuming $\partial_t \rho_{ei} \approx 0$ for $i=0,1,a,e$, resulting in a system of linear equations
\begin{equation}
	\left\{ 
	\begin{aligned}
		& - \Gamma_e \rho_{e0} + i (J_x + i J_y) \rho_{e1} + \frac{i\Omega}{2} \rho_{ea} \approx  i J_L \rho_{10},
		\\&  i(J_x - i J_y) \rho_{e0} - (\Gamma_e + 2 i J_z) \rho_{e1} + i J_z \rho_{ee} \approx  i J_L \rho_{11},
		\\&  \frac{i\Omega}{2} \rho_{e0} - (\Gamma_e + i \delta) \rho_{ea} \approx   i J_L \rho_{1a},
		\\&  -2 \Gamma_e \rho_{ee} - i J_L \rho_{1e} + i J_L \rho_{e1} \approx 0,
	\end{aligned}
	\right.
\end{equation}
with four variables $\rho_{e0}$, $\rho_{e1}$, $\rho_{ea}$ and $\rho_{ee}$. 
Given that $\Gamma_e \gg J_x, J_y, J_z, J_L, \Omega$, the above linear system can be further approximated by
\begin{equation}
	\left\{ 
	\begin{aligned}
		&    - \Gamma_e \rho_{e0} \approx  i J_L \rho_{10},
		\\&  - \Gamma_e \rho_{e1} \approx  i J_L \rho_{11},
		\\&  - \Gamma_e \rho_{ea} \approx  i J_L \rho_{1a},
		\\&  -2 \Gamma_e \rho_{ee} - i J_L \rho_{1e} + i J_L \rho_{e1} \approx 0,
	\end{aligned}
	\right.
\end{equation}
whose solution is given by
\begin{equation}
	\label{AdiabaticEliminationResults}
	\left\{ 
	\begin{aligned}
		&      \rho_{e0} \approx - i (J_L/\Gamma_e) \rho_{10},
		\\&    \rho_{e1} \approx - i (J_L/\Gamma_e) \rho_{11},
		\\&    \rho_{ea} \approx - i (J_L/\Gamma_e) \rho_{1a},
		\\&  \rho_{ee}  \approx  (J_L^2/\Gamma_e^2) \rho_{11}.
	\end{aligned}
	\right.
\end{equation}
Eq.~(\ref{AdiabaticEliminationResults}) can be used to eliminate the $\ket{e}$ level in Eq.~(\ref{fullME}).

\begin{figure}
	\centering
	\includegraphics[width=0.4\textwidth]{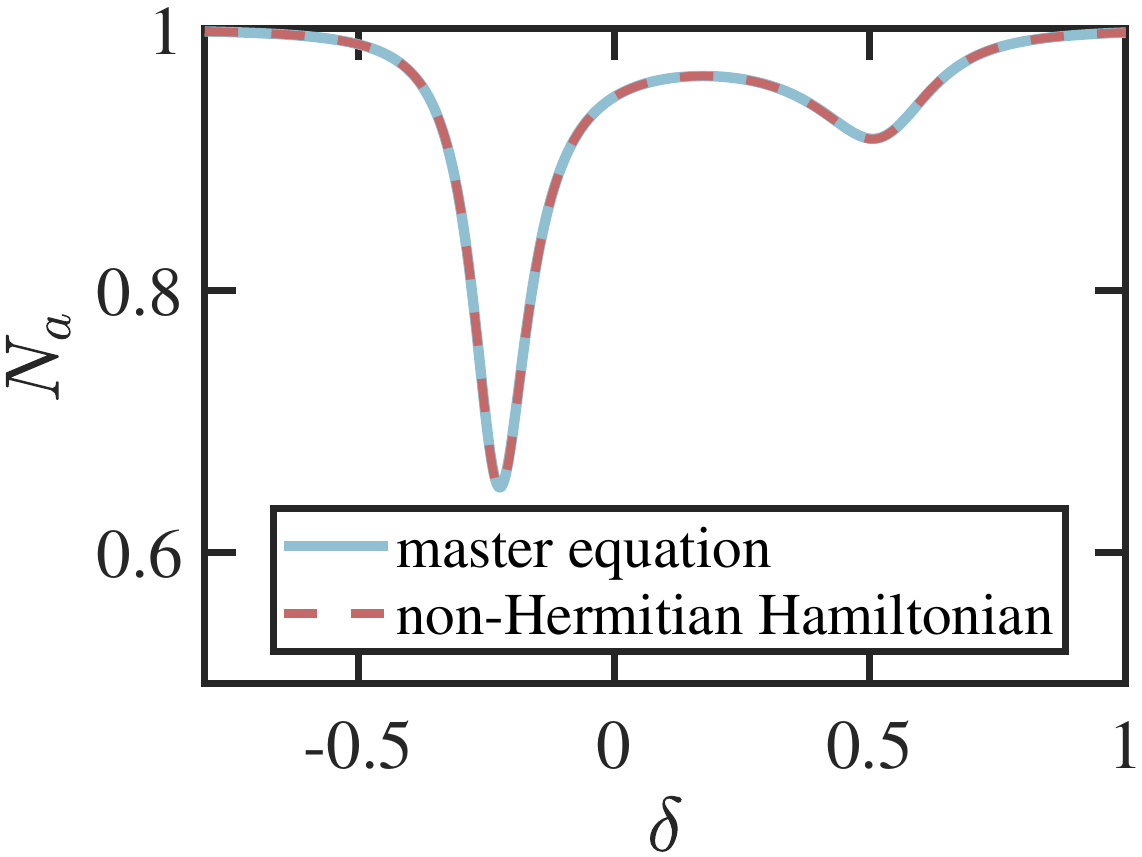}
	\caption{The spectral lines calculated using the master equation (\ref{fullME}) (solid blue line) and the non-Hermitian Hamiltonian in Eq.~(\ref{EffHam}) (dashed red line).
		Two lines coincide with each other, implying that the dynamics of the master equation is very well characterized by the dynamics of the non-Hermitian Hamiltonian.
		The system parameters we take here are the same as in Fig. 2(a1) in the main text.
		Here, $\Gamma_1=\Gamma_2=\Gamma_3=\Gamma_e/3$ and $J_L = 4.76$ MHz.
	} 
	\label{fig1}
\end{figure}

Now we project the entire system to a subsystem consisting of $\ket{0}$, $\ket{1}$ and $\ket{a}$ levels using the projection operator $P_f = \ket{0}\bra{0} + \ket{1} \bra{1} + \ket{a} \bra{a}$.
We apply the projection operator $P_f$ on Eq.~(\ref{fullME}) and get
\begin{equation}
	\begin{aligned}
		\frac{d \rho_f}{dt} &= -i P_f H_\text{eff} \rho P_f + i P_f \rho H_\text{eff}^\dagger P_f + P_f \sum_{\mu=1}^{3} 2 L_\mu \rho L_\mu^\dagger P_f
		\\ &= -i [H_{f0}, \rho_f]  + [2 \Gamma_1 \rho_{ee} + i J_L (\rho_{1e} - \rho_{e1})]  \ket{1}\bra{1} + i J_L (\rho_{0e} \ket{0}\bra{1} - \rho_{e0} \ket{1} \bra{0} + \rho_{ae} \ket{a}\bra{1} - \rho_{ea} \ket{1}\bra{a}),
	\end{aligned}
\end{equation}
with $\rho_f = P_f \rho P_f$ and $H_{f0} = P_f H P_f = H_{h} + H_c$.
Using Eq.~(\ref{AdiabaticEliminationResults}), we obtain
\begin{equation}
	\begin{aligned}
		\frac{d \rho_f}{dt} & \approx  -i [H_{f0}, \rho_f]  - 2 (\Gamma_e - \Gamma_1) \frac{J_L^2}{\Gamma_e^2} \rho_{11}  \ket{1}\bra{1} - \frac{J_L^2}{\Gamma_e} (\rho_{01} \ket{0}\bra{1} + \rho_{10} \ket{1} \bra{0} + \rho_{a1} \ket{a}\bra{1} + \rho_{1a} \ket{1}\bra{a})
		\\& \begin{aligned}
			\  = &-i [H_{f0}, \rho_f] - 2 (\Gamma_e - \Gamma_1) \frac{J_L^2}{\Gamma_e^2} \ket{1}\bra{1} \rho_f \ket{1}\bra{1} 
			\\& - \frac{J_L^2}{\Gamma_e} (\ket{0}\bra{0}\rho_f\ket{1}\bra{1} +  \ket{1} \bra{1}\rho_f \ket{0} \bra{0} +  \ket{a} \bra{a}\rho_f \ket{1} \bra{1} +  \ket{1} \bra{1} \rho_f \ket{a} \bra{a})
		\end{aligned}
		\\&= -i [H_{f0}, \rho_f] - 2 (\Gamma_e - \Gamma_1) \frac{J_L^2}{\Gamma_e^2} \ket{1}\bra{1} \rho_f \ket{1}\bra{1} + \frac{J_L^2}{\Gamma_e} ( 2 \ket{1} \bra{1} \rho_f\ket{1}\bra{1} - \rho_f\ket{1}\bra{1}  -  \ket{1} \bra{1} \rho_f )
		\\&= -i [H_{f0}, \rho_f] - \frac{J_L^2}{\Gamma_e} ( \rho_f\ket{1}\bra{1}  +  \ket{1} \bra{1} \rho_f ) + 2  \frac{J_L^2 \Gamma_1}{\Gamma_e^2} \ket{1}\bra{1} \rho_f \ket{1}\bra{1} .
	\end{aligned}
\end{equation}
Let $H_f = H_{f0} - (i J_L^2/ \Gamma_e) \ket{1} \bra{1} $, one can find that
\begin{equation}
	\label{withM}
	\frac{d \rho_f}{dt} = -i H_{f} \rho_f + i \rho_f H_f^\dagger   + 2  \frac{J_L^2 \Gamma_1}{\Gamma_e^2} \rho_{11} \ket{1}\bra{1} .
\end{equation}
Furthermore, the dynamics of $\rho_{11}$,
\begin{equation}
	\frac{d \rho_{11}}{dt} = -i \bra{1} [H_{f0},\rho_f] \ket{1} - 2 (\Gamma_2 + \Gamma_3) \frac{ J_L^2}{\Gamma_e^2}  \rho_{11},
\end{equation}
suggesting that $\rho_{11} \approx 0$ because of the dissipation term $- 2 (\Gamma_2 + \Gamma_3) \frac{ J_L^2}{\Gamma_e^2}  \rho_{11}$ on the right-hand side.
Thus, Eq.~(\ref{withM}) can be further approximated by
\begin{equation}
	\label{withoutM}
	\frac{d \rho_f}{dt} \approx -i H_{f} \rho_f + i \rho_f H_f^\dagger.
\end{equation}
Based on Eq.~(\ref{withoutM}), we conclude that the dynamics of a system including $\ket{0}$, $\ket{1}$ and $\ket{a}$ levels is described by an effective non-Hermitian Hamiltonian, 
\begin{equation}
	\label{EffHam}
	H_f =  J_x \sigma_x + J_y \sigma_y - 2 (J_z + i \gamma) \ket{1} \bra{1} + \frac{\Omega}{2} (\ket{0} \bra{a} + \ket{a} \bra{0}) - \delta \ket{a} \bra{a},
\end{equation}
with the decay rate being $\gamma = J_L^2 / (2 \Gamma_e)$ which can be tuned by varying the laser power controlling the value of $J_L$.
We have also numerically verified that the dynamics of the master equation (\ref{fullME}) is very well characterized by the dynamics based on the effective 
non-Hermitian Hamiltonian in Eq.~(\ref{EffHam}), as illustrated in Fig.~\ref{fig1}.

\section{S-2. Experimental setup} \label{sec1}

In this section, we will show the experimental details about our system.
In our experiment, the $^{171} \text{Yb}^{+}$ ion is confined by electrical fields of a linear Paul trap with segmented blade electrodes. 
We apply a $2\pi \times 22.68$ MHz RF field on two blades to generate pseudo potential in the radial direction, and DC voltages with proper gradient on $5$ pairs of segments in the other two blades to provide confinement in the axial direction. 
The qubits are encoded by $\ket{0} \equiv |^{2}{S}_{1/2}, F=0, m_F=0 \rangle$ and $\ket{1} \equiv |^{2}{S}_{1/2}, F=1, m_F=0 \rangle$ in the $^{171} \text{Yb}^{+}$  ground state manifold with hyperfine splitting $\omega_\text{HF}=12.642812$ GHz. 
One microwave with frequency $\omega_\text{MW1}=\omega_\text{HF}+\Delta_B -\Delta_\text{MW1}$ drives the transition between the two qubit states, where $\Delta_B=310.8 \ B^2$ Hz, $B$ is the magnetic field in unit of Gauss (Gs) and $\Delta_\text{MW1}$ is the detuning. 
Another microwave with frequency $\omega_\text{MW2} = \omega_\text{HF}+\Delta_B-\Delta_\text{MW2}$ is used to compensate the AC Stark shift. 
Each microwave frequency is generated by mixing one channel of a $1$ GS/s arbitrary wave generator (AWG) with a stable $12.4$ GHz signal source. 
All signal generators are carefully synchronized to a Rubidium clock with $10$ MHz reference signal using equal length of wires.
We use a pair of Helmholtz coils to create a magnetic field which is aligned perpendicular to the light path and the surface. 
As a result, we get a magnetic field around $8.5$ Gs and the corresponding Zeeman splitting of $\ket{^2 {S}_{1/2}, F=1}$ levels is approximately $12$ MHz. 

A $369$ nm laser beam is used to cool the ion with the aid of $14.7$ GHz electro-optic modulator (EOM) sideband, optically pump the ion to the ground state $\ket{0}$ using a $2.11$ GHz EOM sideband, and detect the state of the ion after turning off all sidebands~\cite{Bruzewicz2019}. 
We also use a $935$ nm laser beam with a $3.07$ GHz EOM sideband to repump the leakage to $^{2}D_{3/2}$ metastable levels back to the Doppler cooling cycle. 
For detection, the $369$ nm laser beam containing both $\sigma_{\pm}$ and $\pi$ polarized components is shined on the ion to excite all Zeeman levels in $|^2S_{1/2}, F=1\rangle$ to the excited state $\ket{e} = |^2P_{1/2}, F=0, m_F = 0\rangle$, which will decay in several nanoseconds. 
Fluorescence photons generated by spontaneous emission from the $|e\rangle$ level will be collected by a homemade object with NA=$0.13$, 
and then imaged by a photon-multiplier tube (PMT) controlled by a field-programmable gate array (FPGA). 
We use a threshold method to identify the state of an ion~\cite{Wineland1998,Zhao2022CP}. 
For a fixed detection time, if the measured photon count exceeds the detection threshold, the ion is identified as occupying the bright states in the $|^2S_{1/2},F=1\rangle$ manifold; otherwise, the ion is identified as occupying the dark state $|0\rangle$. 
Experimentally, we set the detection time to $400$ $\mu$s and the detection threshold to $1$, which leads to a detection fidelity of $98.0\%$ for the bright state.
Similiar to the detection of ion states, population on the auxiliary level $|a\rangle$ is detected based on this protocol with different definition of bright and dark states.
During the detection of $N_a$, the $14.7$ GHz EOM of the $369$ nm laser is turned on, which enables all transitions from $^2S_{1/2}$ to $^2P_{1/2}$, so that all states on $^2S_{1/2}$ are bright states. At the same time, the $3.07$ GHz EOM of 935nm repumping beam is turned off so that the state $\ket{a}$ cannot be repumped back, making it a dark state. 
As a result, we achieve a bright state detection fidelity up to $99.5\%$ using the same detection time and threshold.

The dissipation on the state $\ket{1}$ is realized by another 369 nm laser beam that only excites the $|1\rangle$ level to the excited state $|e\rangle$. 
Filtered by a Glan-Taylor polarizer, this beam contains only $\pi$-polarized components, 
such that excitations from other Zeeman states ($|^2S_{1/2}, F=1, m_F = \pm1\rangle$) are blocked by selection rules.
The unstable excited state $|e\rangle$ will spontaneously decay to the Zeeman states on $|^2S_{1/2},F=1\rangle$ with equal probabilities. Consequently, there is a net loss of population on $|1\rangle$ whose rate is determined by the laser intensity.

A tilted beam of $435$ nm laser is used to couple the system level  $|0\rangle$ with the auxiliary level $|a\rangle$ by a quadrupole transition and also used for sideband cooling. 
To effectively drive this transition with a narrow linewidth ($3.02$ Hz), we generate the $435$ nm laser by doubling the 871 nm seed laser locked to a high-finesse ULE stable cavity by the PDH scheme~\cite{drever1983}. 
As a result, the linewidth of the $435$ nm laser is $1.27$ kHz estimated from the full width at half maximum (FWHM) of the clock resonance 
($\ket{^2S_{1/2}, F=0, m_F=0} \rightarrow \ket{^{2} D_{3/2},F=2,m_F=0}$) spectrum. This is the upper bound of the actual linewidth owing to the stray AC magnetic field and drift of the cavity. 
In addition,  other lasers are stabilized to wavelength meter by a homemade program which can limit the long time frequency drift within $2$ MHz.

In experiments, we first prepare the ion on the motional ground state by sideband cooling using the $435$ nm laser~\cite{osti_1595883}. The cooling beam is modulated by an acousto-optic modulator (AOM) using double-pass configuration, and thus the intensity and frequency components can be fine-tuned by changing the input signal. 
After $2$ ms Doppler cooling, we optically pump the ion to the ground state $|0\rangle$, and then tune the $435$ nm laser frequency to the red sideband in radial direction $\omega_x= 2.82$  MHz and $\omega_y=3.07$ MHz in turn. 
As a result, we reach the phonon number $\bar{n}_x<0.11$ and $\bar{n}_y < 0.15$ after $50$ rounds ($2$ ms) of sideband cooling.

The experimental sequence is controlled by FPGA running at $50$ MHz. 
All AOMs and EOMs can be switched in microseconds by the TTL signal generated from FPGA.
For one round of experiment, a typical experimental sequence is shown in Fig. 1(c) in the main text. 
We first cool the ion to the ground state by $2$ ms Doppler cooling followed by $2$ ms sideband cooling, and then prepare the ion on the auxiliary level $|a\rangle$ by a 10 $\mu$s $\pi$ flip using the $435$ nm beam. 
Next, we turn on the system Hamiltonian and $435$ nm weak coupling beam with detuning simultaneously.
After holding for $200$ $\mu$s of time evolution, we take $400$ $\mu$s to detect the state of the ion, which corresponds to one shot in the detection of the population $N_a$ on the auxiliary level.

\begin{figure}
	\centering
	\includegraphics[width=14cm]{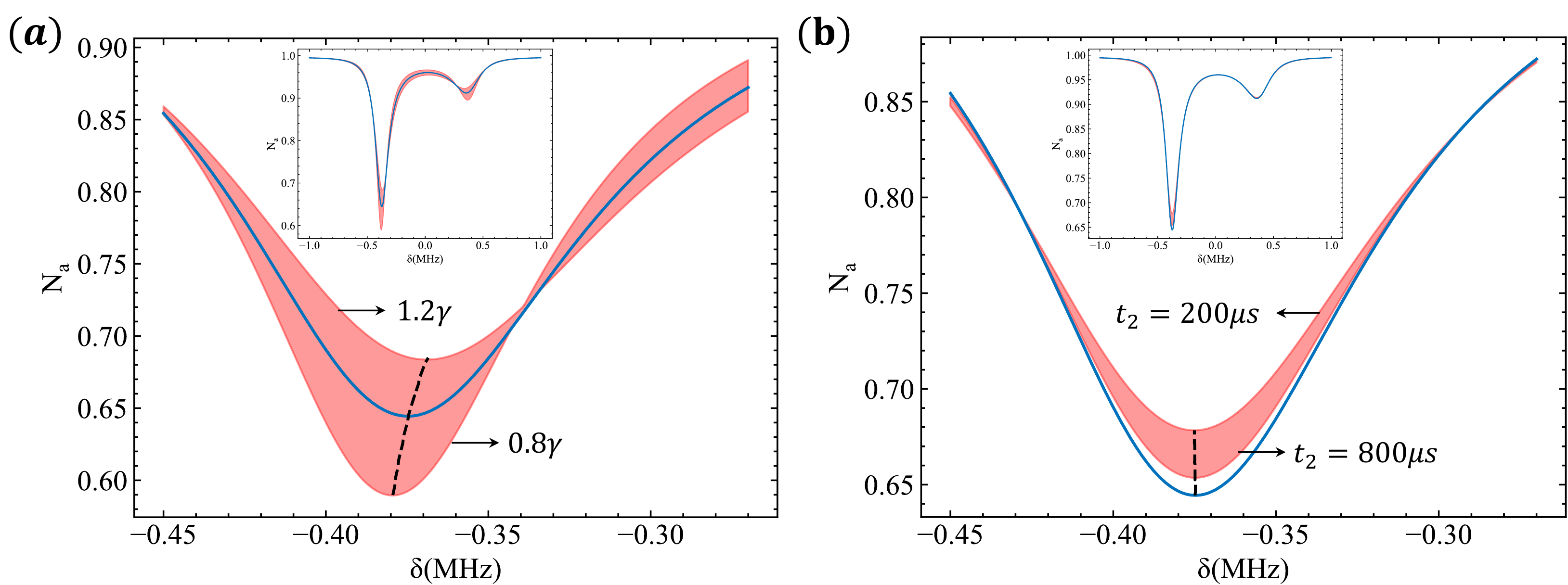}
	\caption{Numerical simulations about error sources. 
		(a) The effect of fluctuations in the intensity of the dissipation laser on the spectral lines.
		The boundaries of the filled region correspond to dissipation rate $\gamma_1 = 0.8\gamma$ and $\gamma_2 = 1.2\gamma$ with $\gamma=0.092$ MHz. 
		(b) Errors induced by the dephasing of the $435$ nm laser beam.
		The filled region is obtained by scanning $t_2$ from $200$ $\mu$s to $800$ $\mu$s. 
		In (a) and (b), the insets represent the spectral lines in a wider range of detuning, the blue solid lines are the theoretical spectral lines under the same parameters as Fig.~2(a1) in the main text, and the black dashed lines corresponds to locations of the dip center.
	}
	\label{fig2}
\end{figure}

\section{S-3. Error Analysis}
In this section, we will provide more analysis about errors in our experiments. 
In one round of experiment, the parameters are relatively stable in our system, and thus we mainly focus on the state preparation and measurement errors and quantum projection noises.
The initial preparation of the ion on the auxiliary level is realized by transferring the state from $\ket{0}$ to $\ket{a}$ with a $\pi$-pulse using the $435$ nm laser. 
However, this initialization is not perfect due to the residual phonon and fluctuations of the laser frequency. 
In average, the preparation fidelity is $99.2\%$ which can be further improved by better sideband cooling and laser locking. 
The detection of the population $N_a$ on the auxiliary level utilizes the threshold method, and reaches an average fidelity of $99.5\%$ when the detection time is set to $400$ $\mu$s;
such a fidelity can be improved using a higher NA object.
And the quantum projection noises can be suppressed by increasing the number of measurements. 
Therefore, we repeat the measurement of the final state for $1000$ times to acquire the expectation value.

However, during many rounds of experiments, the fluctuations of experimental parameters such as microwave or laser intensity, especially the intensity of the dissipation light, are the dominant errors, which cannot be simply eliminated by increasing the number of measurements. 
From Eq.~(\ref{EffHam}), one can find that fluctuations of the dissipation light will lead to the variation of $\gamma$, which introduces errors to the measurement of complex eigenenergies.
To demonstrate the effect of intensity fluctuations on the measurement results, we have numerically simulated the evolution of $N_a$ of the {modified} 
non-Hermitian Rice-Mele model under different dissipation rates.  
From the results shown in Fig.~\ref{fig2}(a), we find that as $\gamma$ increases, the absorption dip gets wider and shallower and the position of the dip center also moves slightly, indicating that fluctuations of the dissipation light will lead to errors in both the real and the imaginary parts of eigenenergies. 

Besides, the decoherence between the system level $\ket{0}$ and the auxiliary level $\ket{a}$ caused by the phase noise of the $435$ nm laser cannot be ignored. 
In our experiment, the typical dephasing time $t_2$ of a well locked $435$ nm laser ranges from $400$ $\mu$s to $800$ $\mu$s measured by the Ramsey fringes.
We have done a numerical simulation by considering the effect of laser dephasing in the master equation.
From the simulation results in Fig.~\ref{fig2}(b), one can find that when dephasing happens, the dip becomes shallower and wider while its location is unchanged, 
indicating that errors caused by the laser dephasing mainly reside in the imaginary parts of eigenenergies.

\section{S-4. Detecting the complex energy spectra with a shorter evolution time}

\begin{figure}
	\centering
	\includegraphics[width=8cm]{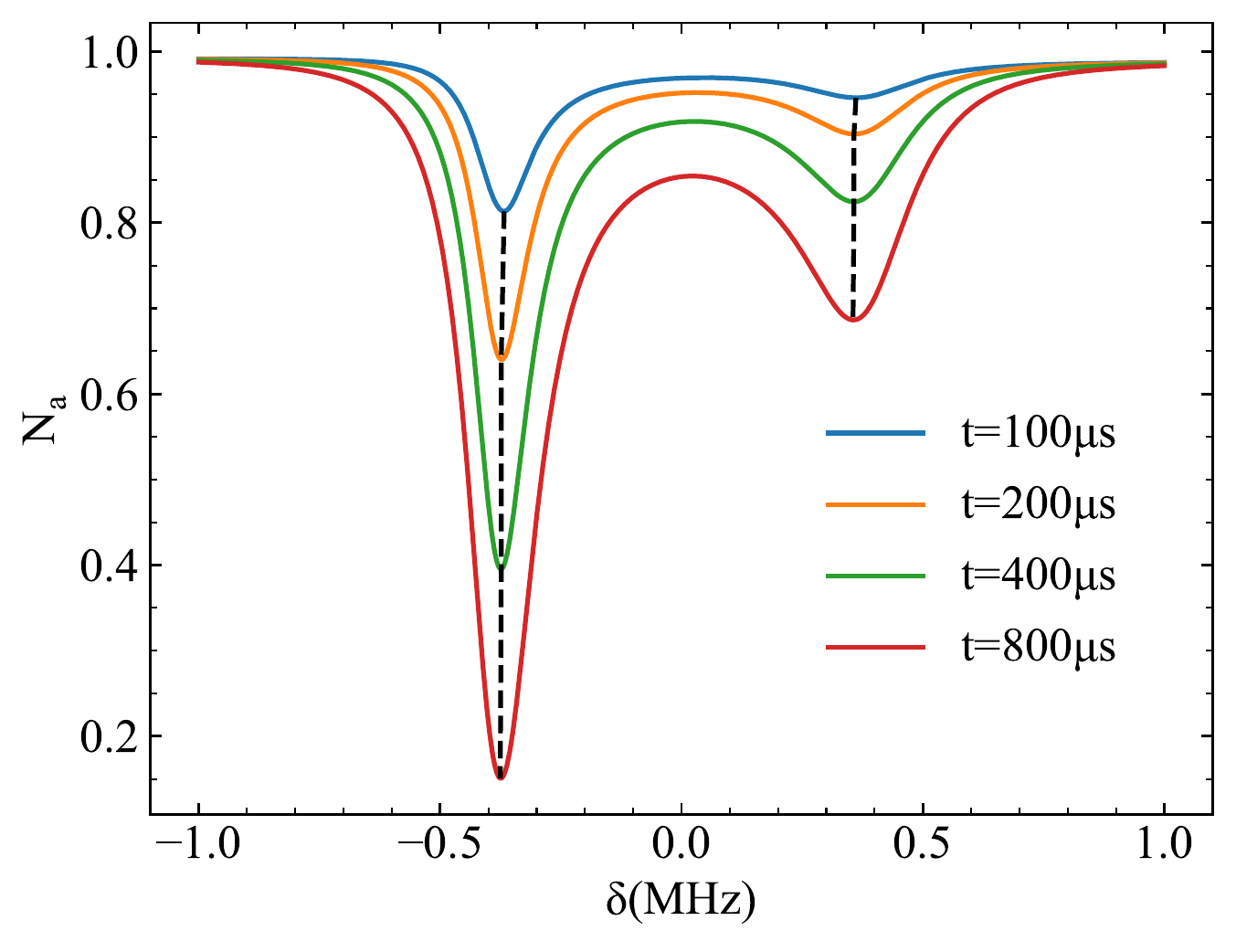}
	\caption{Numerically simulated spectral lines under different evolution time $t$, with other parameters being the same as Fig.~2(a1) in the main text.
		The position of the dip centers are marked by black dashed lines.
	}
	\label{fig3}
\end{figure}

In this section, we will demonstrate the feasibility of detecting complex energies using a shorter evolution time.
To start with, we numerically simulate the spectral lines under different evolution times as shown in Fig.~\ref{fig3}.
We see that as evolution time increases, the spectral line gets sharper, making it easier to locate the position and half width of the absorption dip.
In contrast, for the shortest evolution time $t=100$ $\mu$s (blue solid line), a shallow absorption dip makes the information about the eigenenergy within the dip hard to be extracted, especially in the presence of experimental noises.
The results suggest that a longer evolution time may be beneficial for extracting the complex eigenenergies. 
In the main text, we have set the evolution time to $200$ $\mu$s considering the laser dephasing and fluctuations of system parameters.

However, it is yet a challenge for some systems to keep coherent and stable for a long evolution time. 
To demonstrate the feasibility of detecting complex energies in those systems, we adopt a shorter evolution time in the detection of complex eigenenergies.
We plot the experimentally extracted complex spectra of the {modified} non-Hermitian Rice-Mele model in Fig.~\ref{fig4}, with the same system parameters as in Fig.~2(a1-d1) in the main text except that the evolution time is $t=80$ $\mu$s.
Compared with results in Fig.~2(a1-d1) in the main text, 
the errors in the imaginary parts, especially for those with greater $|\mathrm{Im}(E)|$, get larger as shown in Fig. \ref{fig4}(d).
This can be attributed to the fact that a smaller evolution time makes the dip shallower, and thus the information about imaginary parts become more sensitive to experimental noises.
In spite of larger errors in the imaginary parts, the experimental results still agree with theoretical ones well and clearly captures the loop structure of the energy spectra in the complex-plane, which demonstrates the feasibility of detecting the complex energy spectra in a short time. 

\begin{figure}
	\centering
	\includegraphics[width=\textwidth]{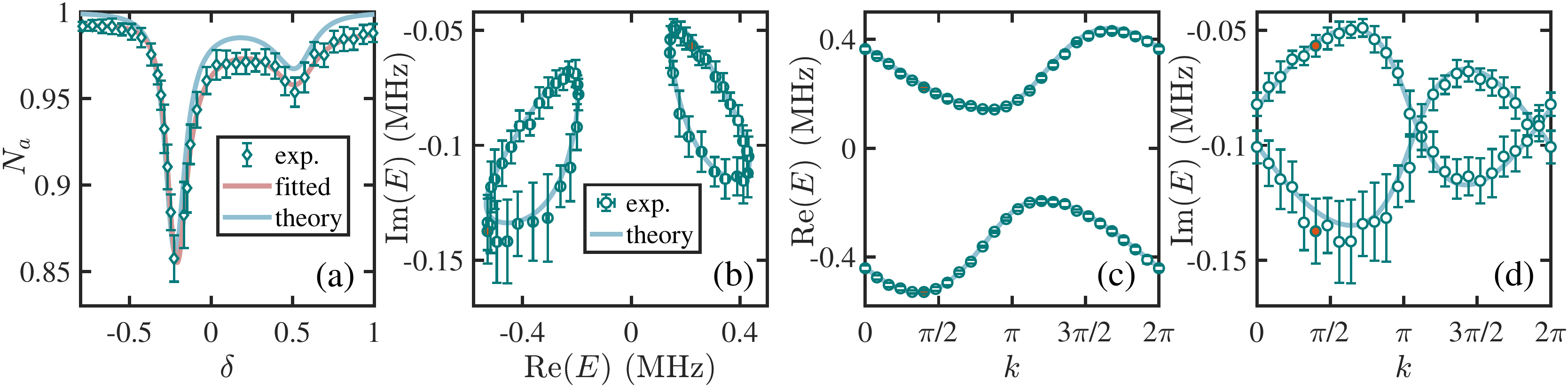}
	\caption{Experimentally measured complex energies of the {modified} non-Hermitian Rice-Mele model. 
		The parameters are taken the same as in Fig. 2(a1-d1) in the main text except that the evolution time $t$ is $80$ $\mu$s. 
		(a) The spectral line obtained by scanning the detunning when the momentum $k=2\pi/5$, which is further used to extract the complex eigenenergies [the corresponding extracted energies are highlighted by red circles in (b)–(d)].
		(b) Complex eigenenergies in the complex-energy plane. 
		The real (c) and imaginary (d) parts of experimentally measured complex energies. 
	}
	\label{fig4}
\end{figure}


\end{widetext}

\end{document}